\documentclass[%
 reprint,
 showpacs,
 amsmath,amssymb,
 aps,
]{revtex4-1}
\usepackage{amsmath}
\usepackage{cases}
\usepackage{amssymb}
\usepackage{CJK}
\usepackage{graphicx}
\usepackage{dcolumn}
\usepackage{bm}
\usepackage{color}
\usepackage[pagewise]{lineno}

\begin{document}

\preprint{APS/123-QED}

\title{Temperature determined by isobaric yield ratio in heavy-ion collisions}

\author{C. W. Ma$^{1}$}\thanks{Email: machunwang@126.com}
\author{J. Pu$^{1,2}$}
\author{Y. G. Ma$^{2}$}\thanks{Email: ygma@sinap.ac.cn}
\author{R. Wada$^{3}$}\thanks{Email: wada@comp.tamu.edu}
\author{S. S. Wang$^{1}$}

\affiliation{$^{1}$ Department of Physics, Henan Normal University, Xinxiang 453007, China\\
$^{2}$ Shanghai Institute of Applied Physics, Chinese Academy of Sciences, Shanghai 201800, China\\
$^{3}$ Institute of Modern Physics HIRFL, Chinese Academy of Sciences, Lanzhou 730000, China
}

\date{\today}

\begin{abstract}
\begin{description}
\item[Background] Temperature ($T$) in heavy-ion collision is an important
parameter. Previously, many works have focused on the
temperature of the hot emitting source. But there are few systematic
studies of the temperature among heavy fragments in peripheral collisions with
incident energies near the Fermi energy to a few $A$ GeV, though it is very
important to study the property of neutron-rich nucleus in heavy-ion
collisions.
\item[Purpose] This work focuses on the study of temperature associated
with the final heavy fragments in reactions induced by both the neutron-proton
symmetric and the neutron-rich projectiles, and with incident energy
ranges from 60$A$ MeV to 1$A$ GeV.
\item[Methods] Isobaric yield ratio (IYR) is used to determine
the temperature of heavy fragments. Cross sections of measured
fragment in reactions are analyzed, and a modified statistical
abrasion-ablation (SAA) model is used to calculate the yield of
fragment in 140$A$ MeV $^{64}$Ni + $^{9}$Be and 1$A$ GeV $^{136}$Xe + $^{208}$Pb reactions.
\item[Results] Relatively low $T$ of heavy fragments are obtained
in different reactions ($T$ ranges from 1 to 3MeV). $T$ is also
found to depend on the neutron-richness of the projectile. The incident
energy affects $T$ very little. $\Delta\mu/T$ (the ratio of the
difference between the chemical potential of neutron and proton to
temperature) is found to increase linearly as $N/Z$ of projectile
increases. It is found that $T$ of the $^{48}$Ca reaction, for which IYRs are of
$A<50$ isobars, is affected greatly by the temperature-corrected $\Delta B(T)$. But $T$ of reactions using IYRs of heavier fragments are only
slightly affected by the temperature-corrected $\Delta B(T)$. The SAA
model analysis gives a consistent overview of the results extracted
in this work.
\item[Conclusions] $T$ from IYR, which is for secondary fragment, is
different from that of the hot emitting source. $T$ and $\Delta\mu$
are essentially governed by the sequential decay process.
\end{description}
\end{abstract}

 
\maketitle

\linenumbers

\section{introduction}

In heavy-ion collisions (HIC) above the intermediate energy
(with incident energy $>20A$ MeV), the temperature ($T$) is high enough to provide the
environment for nuclear liquid-gas transition. Many works have
focused on the critical point in $T$ experimentally
\cite{ModelFisher3,Goodman84Tc,YgMa99PRCTc,YGMa04NST,EOSPRC02Tc,DasPRC02Tc,YgMa04PRCTc,YgMa05PRCTc}
and theoretically \cite{Glen86PRCTc,Baldo99PRCTc}.
$T$ is also important in determining the symmetry energy
of a neutron-rich nucleus \cite{MBTsPRL01iso,Botv02PRCiso,Ono03PRCiso,Ono04RRCiso,SouzaPRC09isot}.
Due to the complexity of the HIC processes and the importance of
$T$, many methods have been developed to obtain $T$
from thermal energy\cite{Bonasera11PLBT}, excitation energy \cite{ZhouPei11T,Morr84PLBT_E,Poch85PRL_E,Nato02PRCT_E}, isotopic
yields (Albergo thermometer) \cite{YgMa05PRCTc,AlbNCA85DRT,Nato95T,Poch95PRLDR_T,JSWangT05,Wada97T,Surfling98TPRL},
and kinetic energy spectra \cite{Westfall82PLBT_spec,Jacak83T_spec,SuJPRC12T_spec}.
$T$ based on these methods have differences but can be related to each other \cite{Poch85PRL_E,SuJPRC12T_spec}.

The yield of fragment in HIC, on the one hand is greatly influenced by $T$,
and on the other hand, is determined by its free energy due to non-zero
$T$ \cite{ModelFisher3}. Thus the yield of fragment can constrain both its
binding energy and $T$. In some models which estimate isotopic yield,
$T$ is an important parameter \cite{Tsang07BET,Chau07PRCTf,Mallik11PRCFt,Souza12PRCFn},
while in other works the isotopic yield is also used to estimate the
binding energy of unknown isotopes \cite{Tsang07BET}. In works using
the yield of fragment to constrain the binding energy, for example,
in the study of the symmetry energy of fragments, it is proposed
that the isobaric yield ratio cancels out the energy term which
only depends on the mass of fragment, thus the specific energy term can
be extracted \cite{Huang10,MA12CPL09IYRAsbsbv}. But the shortcoming of this
method is that the coefficient of energy-term and $T$
can not be separated, thus they must be viewed as whole parameters,
for example, in the study of the symmetry-energy coefficients to
$T$ ($a_{sym}/T$) of the neutron-rich nucleus \cite{Huang10,MaCW11PRC06IYR,MaCW12EPJA,MaCW12CPL06}.
If the binding energy of fragment at nonzero $T$ is known,
the yield of fragment can also be used to determine $T$. One point
to remember is that the measured heavy fragment in HIC undergoes
the sequential decay and deexcitation processes, which makes it cool
down. $T$ of a heavy fragment should be quite lower than that of the
hot emitting source \cite{DAS02PRCTf,ZhouPei11T}.

Based on the free energy, the modified Fisher model (MFM)
well describes the isotopic yield distributions of intermediate
mass fragments produced in proton-induced multifragmentation at
relativistic energies \cite{ModelFisher3,ModelFisher1}. The MFM
has been used to study the behavior of fragments near the critical
point of the liquid-gas transition \cite{YgMa99PRCTc,EOSPRC02Tc,DasPRC02Tc,YgMa04PRCTc,YgMa05PRCTc}.
In the MFM, the yield of an isotope is determined by chemical potential,
free energy and entropy. In this article, we will use the isobaric
yield ratio to constrain $T$ in HICs. The article
is organized as follows: First, we briefly introduce the isobaric
ratio method based on MFM. Second, we verify that the
binding energy of nucleus at zero temperature can be used instead in the isobaric ratio
method. At last, $T$ determined by isobaric yield will be shown and the
results will be discussed.

\section{Isobaric yield ratio method}

Following the MFM \cite{ModelFisher3,ModelFisher1},
the generalized expression of the yield of a fragment with mass $A$ and
neutron-excess $I (I \equiv N - Z)$ is
\begin{eqnarray}\label{yeild}
Y(A,I) = CA^{-\tau}exp\{[W(A,I)+\mu_{n}N+\mu_{p}Z]/T  \nonumber\\
+Nln(N/A)+Zln(Z/A)\},
\end{eqnarray}
where $C$ is a constant. The $A^{-\tau}$ term originates from the
entropy of the fragment, $\tau$'s for all fragments are
identical. $\mu_n$ and $\mu_p$ are the neutron and proton chemical
potentials, respectively, and $W(A,I)$ is the Helmholtz free energy
of the cluster (fragment). In principle Eq. (\ref{yeild})
should be applied to hot nuclear matter near the critical point.
However, when Eq. (\ref{yeild}) was applied to the cold fragments
in Ref. \cite{Huang10PRCTf}, they showed that useful information can
be extracted to elucidate the effects of the secondary decay process.
In that case, $T$ and other parameters do not correspond to those in
the primary hot nuclear matter, but do correspond to those modified by the secondary
decay process. Therefore it is still useful to apply Eq. (\ref{yeild})
to the experimentally observed cold fragments to elucidate the effect
of the sequential process on the characteristic physical parameters,
such as $T$ or chemical potential, as discussed below.

Defining the yield ratio between isobars differing by 2 units in $I$, we have
\begin{eqnarray}\label{ratiodef}
&R(I+2,I,A)
=Y(A,I+2)/Y(A,I)  \nonumber\\
& =\mbox{exp}\{[W(I+2,A)-W(I,A)+\Delta\mu]/T \nonumber\\
& +S_{mix}(I+2,A)-S_{mix}(I,A)\},
\end{eqnarray}
where $S_{\rm mix}(I,A)=N\mbox{ln}(N/A)+Z\mbox{ln}(Z/A)$, and $\Delta\mu=\mu_n-\mu_p$.
Taking the logarithm of $R(I+2,I,A)$, one obtains
\begin{equation}\label{tlnRcal}
\mbox{ln}R(I+2,I,A)-\Delta S=(\Delta W+\Delta\mu)/T
\end{equation}
where $\Delta S = S_{\rm mix}(I+2,A)-S_{\rm mix}(I,A)$,
and $\Delta W = W(I+2,A)-W(I,A)$ is the difference between
the free energies of isobars. $W(I,A)$ is supposed to equal
the binding energy ($B$) of a fragment at given $T$
and density $\rho$ \cite{ModelFisher3} [written as $B(\rho,T)$].
If $B(\rho,T)$ (which includes the contributions from
the binding energy and entropy) is known, $T$ and $\Delta\mu$
can be obtained using the isobaric yield ratio (IYR) from Eq. (\ref{tlnRcal}).

The known $B$ of the nucleus is for $T=0$ [written as $B(0)$]. When
$T\neq0$, entropy contributes to the binding energy
[written as $B(T)$] and makes $B(0)\neq B(T)$. The other good
news is that since $\Delta W$ can serve as the independent variable,
actually $B(T)$ is not important anymore in Eq. (\ref{tlnRcal}).
If $B(0)$ can be used instead of $B(T)$ in the calculation of
$\Delta W$, the temperature of the fragment can be extracted using
Eq. (\ref{tlnRcal}). Thus before determining $T$ using the IYR
method, whether $\Delta B(0)=\Delta B(T)$ should be evaluated
[$\Delta B(T)=B_{(I+2)}(T)-B_{(I)}(T)$ is the difference between
the binding energies of the $I+2$ and $I$ isobars].

Using the density-functional theory based on the Skyrme interaction (SKM),
the $T$ dependence of the binding energy of a finite nucleus has been
proposed to be \cite{LeeBE_T10PRC},
\begin{eqnarray}\label{BETdep}
B(A,I,T)=&-(15.31-0.04T^2)A~~~~~~~~~~~~~~~~~~~~~~  \nonumber\\
&+(18.30+0.50T^2)A^{2/3} ~~~~~~~~~~~~~~~~~~ \nonumber\\
&+(19.69+0.42T^2)I^2/A ~~~~~~~~~~~~~~~~~~ \nonumber\\
&-(33.18+2.06T^2)I^2/A^{4/3}+E_c\frac{Z^2}{A^{1/3}} \nonumber\\
&+E_{dif}\frac{Z^2}{A}+E_{ex}\frac{Z^{4/3}}{A^{1/3}}+\Delta(N,Z),
\end{eqnarray}
where $E_{dif}$ and $E_{ex}$ are the coefficients for the diffuseness
correction and the exchange correlation to the Coulomb energy.
$\Delta(N,Z)$ is the pairing-energy term.

\begin{figure}[htbp]
\includegraphics
[width=8.6cm]{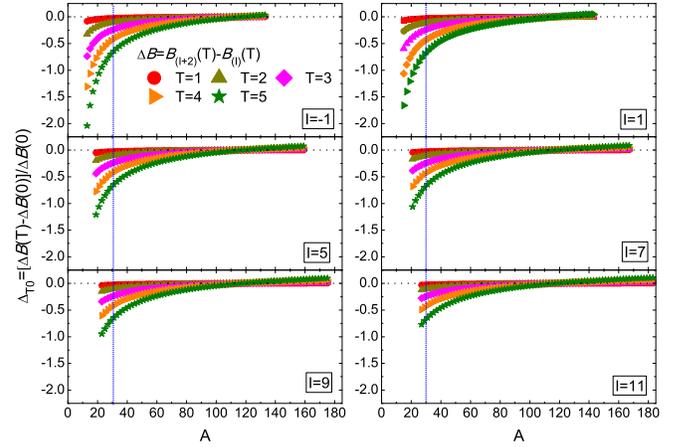}\caption{\label{BET1-6} (Color
online) $\Delta_{T0}\equiv[\Delta B(T)-\Delta B(0)]/\Delta B(0)$ of isobars with $I$ from -1 to 11 at
different $T$. $\Delta B(T)$ is the difference between the binding
energies of the $I+2$ and $I$ isobars at $T$. $\Delta B(T)=B_{(I+2)}(T)-B_{(I)}(T)$.}
\end{figure}

The binding energy of the nucleus at $T$ from 0 to 5MeV are calculated
using Eq. (\ref{BETdep}). To see how fast $\Delta B(T)$ between isobars
increases with temperature, the values of
$\Delta_{T0}\equiv[\Delta B(T)-\Delta B(0)]/\Delta B(0)$ are plotted
in Fig. \ref{BET1-6}. At $T\sim1$MeV, $\Delta_{T0}$ of the $I=-1\sim11$
isobars are very close to zero. $\Delta_{T0}$ decreases as the mass
becomes larger, i.e., the larger the A of nucleus, the closer $\Delta_{T0}$ is to 0.
For isobars of small mass, $\Delta_{T0}$ decreases similarly as $T$
decreases. At $T\geq3$MeV, the $A<50$ isobars show relatively large
$\Delta_{T0}$ ($\Delta_{T0}<-0.1$ corresponds to uncertainty larger
than 10\%). $\Delta B(T)$ of $A<50$ isobars should be used more
carefully around $T\geq3$MeV. In Eq. (\ref{tlnRcal}) $\Delta W$ is
the difference between free energies of isobars, thus the very little
$\Delta_{T0}$ for $I\geq5$ and $A>50$ isobars occurs when $\Delta B(T)$
is replaced by $\Delta B(0)$, and we need not know the actual
$B(T)$. The smaller $\Delta_{T0}$ of isobars at $T$, the closer $T$
obtained from Eq. (\ref{tlnRcal}) approximates the real value. Thus
theoretically, the yield ratios of $I\geq5$ isobars are suitable
observables to extract $T$. Replacing the $\Delta W$ term in
Eq. (\ref{tlnRcal}) by $\Delta B(0)$ (hereafter denoted as $\Delta B$),
one obtains,
\begin{equation}\label{tlnRcalBE}
\mbox{ln}R(I+2,I,A)-\Delta S=(\Delta B+\Delta\mu)/T
\end{equation}

\begin{figure}[htbp]
\includegraphics
[width=8cm]{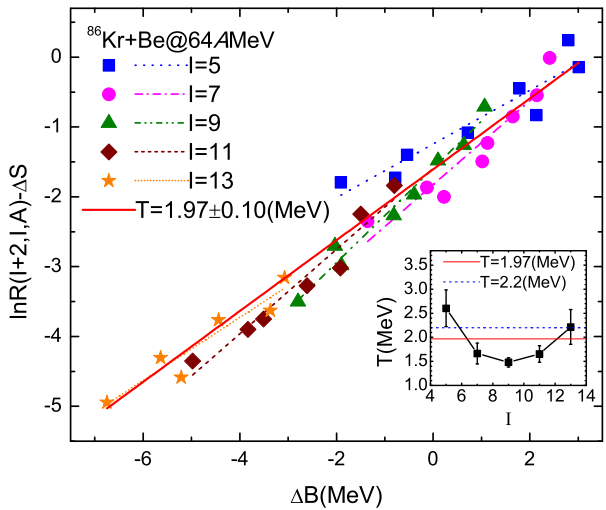}\caption{\label{64AMeVKr86} (Color
online) The correlation between IYR and $\Delta B$ of isobars
in the 64$A$ MeV $^{86}$Kr + $^9$Be projectile fragmentation
reaction \cite{Mocko06KrBe9}. The lines are the fitting results
of IYRs using Eq. (\ref{tlnRcalBE}). The solid line represents
the fitting line of all the data. The inserted figure shows the
fitted values of $T$, in which the solid line represents $T$ fitted
from all the data, and the dotted line represents $T=$2.2 MeV used in
Ref. \cite{Tsang07BET}.
}
\end{figure}

\section{Results and discussions}

By analyzing IYR in HIC, $T$ and $\Delta\mu$ can be determined using
Eq. (\ref{tlnRcalBE}). The analysis is performed using the IYR of
$I\geq5$ isobars. First, the dependence of $T$ on $I$ is investigated
using the data in the 64$A$ MeV $^{86}$Kr + $^9$Be projectile fragmentation
reaction \cite{Mocko06KrBe9}. IYR of isobars are plotted in Fig. \ref{64AMeVKr86},
which almost overlap except the IYR of $I=$5. $\Delta B$ are calculated
using the experimental binding energy of the nucleus in Ref. \cite{BindAudi03}.
The IYRs are fitted individually using Eq. (\ref{tlnRcalBE}) according to
each $I$, and all the data are also fitted as a whole (shown as the
solid line). The fitted values of $T$ are plotted in the inserted figure.
Relatively similar $T$ are obtained from $I\geq5$ IYR. The line in the
inserted figure represents $T$ fitted from all the IYR data.

\begin{figure}[htbp]
\includegraphics
[width=8.0cm]{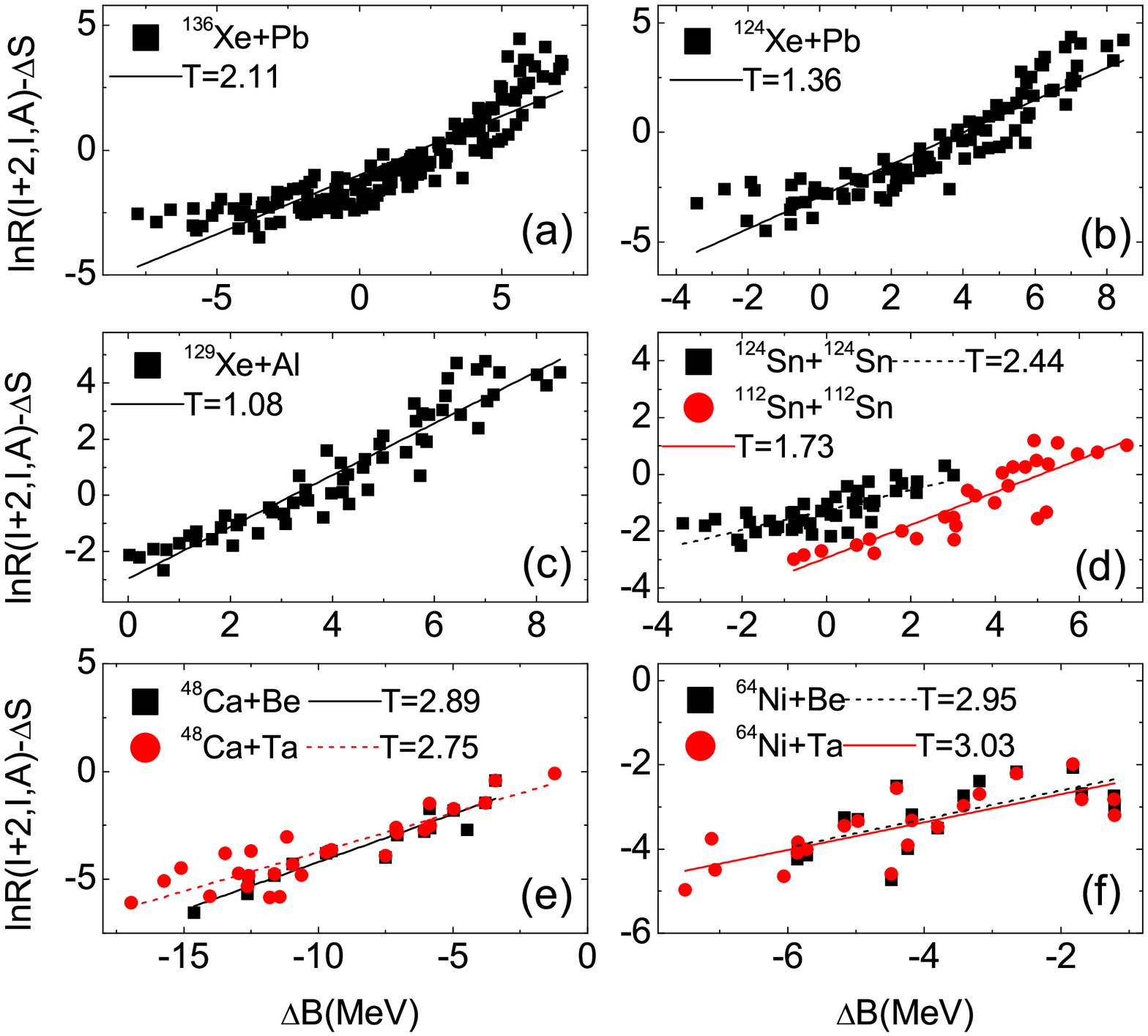}
\caption{\label{lnRTtotal}
(Color online) The correlation between IYR and
$\Delta B$ of isobars in the following reactions:
(a) 1$A$ GeV $^{136}$Xe + Pb \cite{Henz08},
(b) 1$A$ GeV $^{124}$Xe + Pb \cite{Henz08},
(c) 790$A$ MeV $^{129}$Xe + Al \cite{Reinhold98},
(d) 1$A$ GeV $^{112}$Sn+$^{112}$Sn and $^{124}$Sn+$^{124}$Sn \cite{Sn112124},
(e) 140$A$ MeV $^{48}$Ca + $^{9}$Be/$^{181}$Ta \cite{Mocko06}, and
(f) 140$A$ MeV $^{64}$Ni + $^{9}$Be/$^{181}$Ta \cite{Mocko06}. $\Delta B$ is
calculated using the experimental binding energy of the nucleus in Ref. \cite{BindAudi03}.
The lines denote the fitting results using Eq. (\ref{tlnRcalBE}) and
$T$ is the temperature obtained.
}
\end{figure}

\begin{figure}[htbp]
\includegraphics
[width=8.0cm]{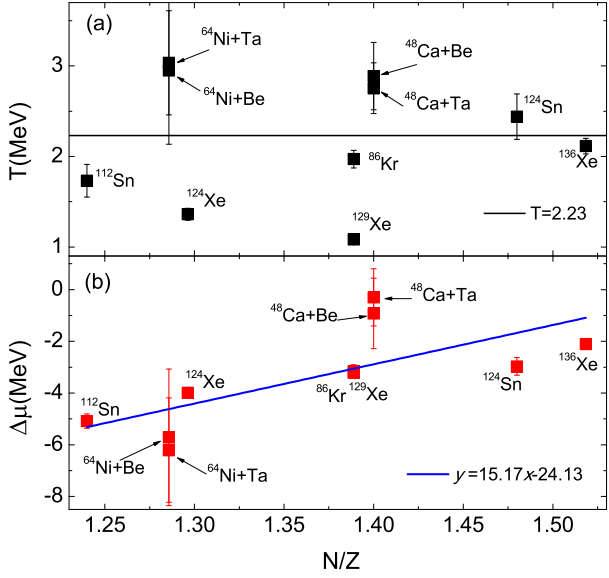}\caption{\label{Tandmu} (Color
online) (a) The fitted temperature ($T$), and (b) $\Delta\mu$ from IYRs
in reactions analyzed in Fig. \ref{lnRTtotal}. $\Delta\mu$ of the $^{86}$Kr
and $^{129}$Xe reactions overlaps. The solid line in (a)
represent the average value of $T$, and the line in (b) is
the result of the linear fit between $\Delta\mu$ and $N/Z$.
}
\end{figure}

To see $T$ extracted from IYRs more systematically, IYRs in reactions of the
1$A$ GeV $^{124,136}$Xe \cite{Henz08}, 790$A$ MeV $^{129}$Xe \cite{Reinhold98},
1$A$ GeV $^{112,124}$Sn \cite{Sn112124}, and 140$A$ MeV $^{48}$Ca and
$^{64}$Ni \cite{Mocko06}, are investigated.
The results are plotted in Fig. \ref{lnRTtotal}. All the IYRs of isobars
with different $I$ are fitted as a whole and $T$ obtained are given
in each panel. IYRs in these reactions can be well fitted using Eq. (\ref{tlnRcalBE}).
$T$ and $\Delta\mu$ obtained are plotted in Fig. \ref{Tandmu}.
The line in Fig. \ref{Tandmu}(a) is the average value of
$T$ in these reactions, which is $T=2.23$MeV.
By analyzing the results of $T$, conclusions below can be drawn:
\begin{itemize}
  \item Relatively low $T$, which ranges from 1MeV to 3MeV, are found in these reactions.
  \item The neutron richness of projectile affects $T$.
  $T$ of a neutron-rich reaction system is higher than that of the
  neutron-proton symmetric reaction systems when similar measurements are made, i.e., $T(^{136}Xe)>T(^{124}Xe)$,
  and $T(^{124}Sn)>T(^{112}Sn)$. This is a similar phenomenon as the isospin dependence
  of the fragment yields measured in reactions of neutron-proton symmetric and neutron-rich
  reactions \cite{MaCW09PRC}. The isotopic temperature ($T_{HeLi}$) was also found
  to increase when the projectile becomes more neutron rich using the isospin-dependent
  quantum molecular dynamics model \cite{SuJun11PRCTiso}.
  \item The mass of target affects $T$ very slightly.
  $T$ obtained from the $^{48}$Ca + $^{9}$Be/$^{181}$Ta
  reaction are very similar. A similar observation in $T$ is made in
  the $^{64}Ni$ + $^{9}$Be/$^{181}$Ta reaction.
  \item The incident energy of the reaction, which ranges from 64$A$ MeV to 1$A$ GeV,
  does not influence $T$ very much. This occurs, as shown later using
  a modified statistical abrasion-ablation  model analysis, because $T$ is essentially governed by the
  sequential decay process.
\end{itemize}

$T$ determined from IYRs of heavy fragments is lower than that
from the Albergo isotopic temperatures of light fragments ($T_{HHe}$)
and slope temperatures ($T_{slope}$) from the energy spectrum \cite{JSWangT05,Poch85PRL_E},
but is close to the isotopic temperatures involving heavier
isotopes ($T_{BeHe}$, $T_{LiBe}$, $T_{BeLi}$, $T_{LiLi}$ \cite{Wada97T}
and $T_{CC}$ \cite{TrautTHFrg07}). In a sequential decay process, light
particles, such as $p, n$, and $\alpha$, are emitted. When the light particle emission ceases,
the fragment can still emit $\gamma$ rays and it further cools off.
The extracted low temperatures associated with heavy
fragments for the reactions in the wide incident energy range,
indicate the dominance of the secondary decay effect on the temperature.
One can also see the systematic trend that the temperatures from the
lighter system tend to be higher than those of the heavier systems.
Even though the MFM method provides a crude way to evaluate the
temperatures, they are still useful in elucidating the entire
process of the fragment production.

$\mu_n,\mu_p$ and $\Delta\mu$ are associated with the properties of the emitting
source but not the fragments themselves \cite{ModelFisher3,ModelFisher1}.
Relatively large $\Delta\mu$ were observed in this analysis for the reactions.
A linear correlation is found between the correlation of
$\Delta\mu$ and $N/Z$ of the projectile, which is shown as the solid line
in Fig. \ref{Tandmu}(b). The linear fitting result between $\Delta\mu$
and $N/Z$ reads $y=(15.21\pm5.82)x+(-24.16\pm7.97)$.

\begin{figure}[htbp]
\includegraphics
[width=8.0cm]{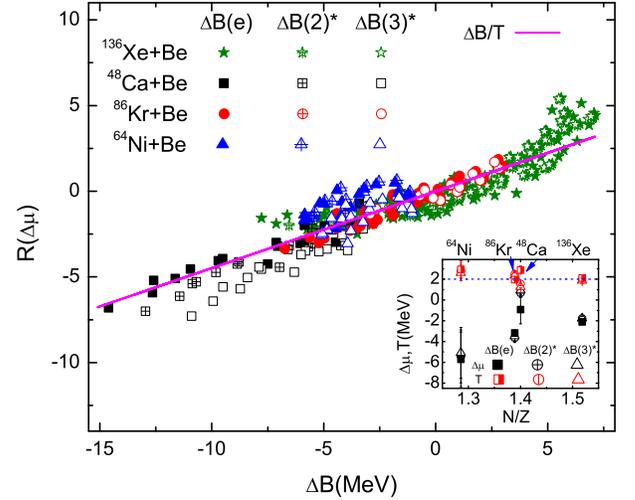}\caption{\label{DBTMod} (Color
online) The correlation between the $R(\Delta\mu)$ and
the temperature-corrected $\Delta B(T)^{*}$ of the $I=(7,5)$ and $I=(9,7)$ IYRs. The solid symbols represent the
results for the experimental $\Delta B(e)$; the crossed and open symbols
represent the results for the temperature-corrected $\Delta B(2)^{*}$ and
$\Delta B(3)^{*}$, respectively. The lines are the fitting
results using Eq. (\ref{tlnRcalBE}). The inserted figure shows the
fitted values of $T$ and $\Delta\mu$ from the original IYR ($ln\mbox{R(I+2,I,A)}-\Delta S$).
}
\end{figure}

Rewriting Eq. (\ref{tlnRcalBE}), we get the following form,
\begin{equation}\label{DBEDivT}
\mbox{ln}R(I+2,I,A)-\Delta S-\Delta\mu/T=\Delta B/T
\end{equation}
with the left hand side involving IYR and $\Delta\mu$, which relate to each
reaction; the right hand side associates with $\Delta B/T$ of the isobars.
For simplification, the left hand side of Eq. (\ref{DBEDivT}) is rewritten as
$R(\Delta\mu)=\mbox{ln}R(I+2,I,A)-\Delta S-\Delta\mu/T$. In Fig. \ref{DBTMod},
$R(\Delta\mu)$ is plotted for typical reactions as a function of $\Delta B$ for
the isobar combinations of  $I=(7,5)$ and $I=(9,7)$. $T$ and $\Delta\mu$ are taken
from Fig. \ref{Tandmu}.  For the $^{64}$Ni reaction, $R(\Delta\mu)$ values are
slightly larger than those predicted by the average $T$. The mass range of
the isotopes with $I=(7,5)$ and $(9,5)$ for $^{64}$Ni reaction is $A=$25 to 63.
It was pointed out earlier that the analysis using $A<50$ isobars has large uncertainty due to large $\Delta_{T0}$
when $T\geq3$MeV. Using the calculation of $\Delta_{T0}$ shown in Fig. \ref{BET1-6},
the experimental $\Delta B(0)$ of isobars are temperature-corrected for
$T=$2 and 3MeV, which are labeled as $\Delta B(T)^{*}$. After the temperature correction,
the correlation between the isobar combinations of $I=(7,5)$ and $I=(9,7)$
and the temperature-corrected $\Delta B(T)^*$ are plotted in Fig. \ref{DBTMod}.
It is easy to see that for the $^{48}$Ca reaction, $T$ from
$\Delta B(2)^{*}$ and $\Delta B(3)^{*}$ decrease to 1.56$\pm$0.12MeV and
1.27$\pm$0.10MeV, respectively. For other reactions, $T$ from the
temperature-corrected $\Delta B(T)^{*}$ are only slightly modified.
The correlation between the $I=(7,5)$ and $I=(9,7)$ IYRs
and $\Delta B(3)^*$ for the $^{64}$Ni reaction overlap with $\Delta B/T$.
It can be concluded that $T$ extracted from IYR of small mass
is greatly affected by $\Delta B(T)^{*}$.

In Refs. \cite{Tsang07BET,Chau07PRCTf,Souza12PRCFn,Das05PR}, different temperatures ($T=$2.2, 6.0 and 9.5MeV)
are used. Comparing to other works, a relatively low $T=$2.2MeV was used to
estimate the binding energies of very neutron-rich copper isotopes \cite{Tsang07BET}
in the 64$A$ MeV $^{86}$Kr + $^9$Be reaction \cite{Mocko06KrBe9}.
The staggering in the isotopic yield is minimized by introducing the
approximation of back-shifted Fermi gas level density, or a parameter $\varepsilon$.
In Fig. \ref{64AMeVKr86}, $T$ and $\Delta\mu$ obtained from all the IYRs is
$T=1.97\pm0.10$MeV and $-3.16\pm0.26$MeV, respectively. The dashed line represents
$T=2.2$MeV. In Ref. \cite{Tsang07BET}, the equation used [Eq. (1)] to
calculate the yield of a fragment $(N,Z)$ is very similar to what we used in this article,
$Y(N,Z)=CA^{3/2}\mbox{exp}[(N\mu_n+Z\mu_p-F)/T]$, where $F$ is the free
energy, $\Delta\mu=-2.5$MeV ($\mu_n=-9.5$MeV and $\mu_p=-7.0$MeV), $\tau=3/2$
is different from the MFM of $\tau=-3.6\sim-2.2$ (minus sign) in
different $I$ values \cite{Huang10PRCTf}. The very similar temperatures
in Ref. \cite{Tsang07BET} and this work indicate that temperature
extracted from isobaric yield ratios is reasonable.

Finally, we investigate the temperature of pre-fragment
and final fragment in a modified statistical abrasion-ablation (SAA)
model \cite{SAABrohm94,FangPRC00}. The SAA model can well reproduce the
yield of fragments \cite{MaCW09PRC} and was used to study the isospin
phenomena in HICs \cite{MaCW10PRC,MaCW09CPB,MACW10JPG,MaCW11CPC}.
In brief, in the SAA model, the pre-fragment is calculated after the numbers of abraded protons
and neutrons are known, which are determined by the nuclear-density distribution
in the overlapping zone of projectile and target, and the nucleon-nucleon
reaction cross section. Mean excitation energy of 13.3$\Delta A$ MeV
is assigned in the initial pre-fragments when $\Delta A$ numbers of
protons and neutrons are removed from the projectile in the
ablation-abrasion process. After the abrasion, the excited initial projectile nucleus undergoes
the deexcitation process and forms the final fragment. The
model description can be found in Refs. \cite{SAABrohm94,FangPRC00,MaCW09PRC}.

\begin{figure}[htbp]
\includegraphics
[width=8.6cm]{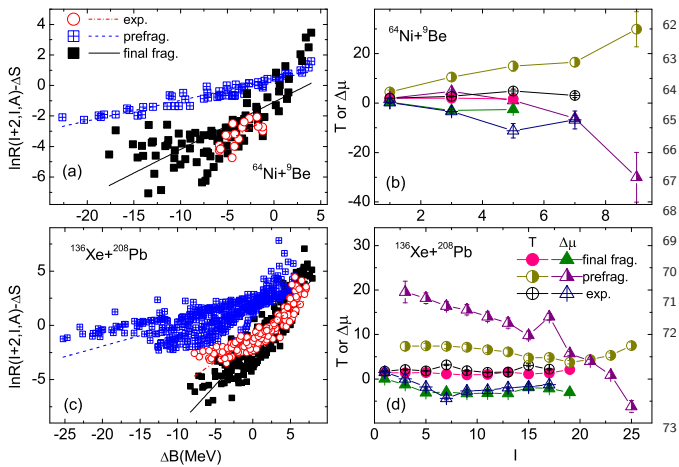}\caption{\label{prefinalSAA} (Color
online) (a) IYRs for the pre-fragments (crossed squares) and final fragments
(solid squares) in the 140$A$ MeV $^{64}$Ni + $^9$Be reaction of
the SAA result, and those for the measured fragments (open circles).
(b) $T$ (circles) and $\Delta\mu$ (triangles) determined from IYRs of
the prefragments (half-full symbols), final fragments (full symbols) and
the measured fragments (crossed symbols) in the 140$A$ MeV $^{64}$Ni + $^9$Be.
(c) and (d) are the same as that of (a) and (b), respectively, but for the
1$A$ GeV $^{136}$Xe + $^{208}$Pb reaction. }
\end{figure}

The 140$A$ MeV $^{64}$Ni + $^9$Be and 1$A$ GeV $^{136}$Xe + $^{208}$Pb reactions
are calculated. The IYRs of pre-fragments and final fragments in the $^{64}$Ni
and $^{136}$Xe reactions are plotted in Fig. \ref{prefinalSAA}(a) and (c),
respectively. The IYRs of final fragments mostly overlap with those of the
measured ones, while the IYRs of the pre-fragments have large difference from
those of final fragments and the measured ones. $T$ and $\Delta\mu$ determined
from the IYRs with different $I$ of pre-fragments and final fragments in the
$^{64}$Ni and $^{136}$Xe reactions are plotted in Figs. \ref{prefinalSAA}(b)
and \ref{prefinalSAA}(d), respectively. $T$ from IYRs of the final fragments are very similar
to those from the experimental fragments, while $T$ from IYRs of the
pre-fragments are rather higher than those from the experimental fragments.
It is shown that the drastic modification of $T$
and chemical potential between the pre-fragments and the final fragments results
in the similar values in $T$ and chemical potential extracted by the
IYRs from the cold fragments in different reactions studied here. In other words,
they are essentially governed by the sequential decay process. Therefore the yield
of the cold fragments can be obtained by a simple scaling of Eq. (\ref{tlnRcalBE})
for a variety of reaction systems in the wide incident energy range studied in this work.
The extracted $T$ plotted in Fig. \ref{Tandmu}(a) show low $T$
values of 1$\sim$3MeV. However, within that range they tend to correlate with the
projectile masses, that is, $T$ decreases as the projectile mass
increases. This may reflect the difference in $T$ of the pre-fragments
before the secondary decays. In the SAA analysis, the excitation energy
of the pre-fragments are given by $E^{*}=13.3\Delta A$ MeV, where $\Delta A$
is the number of nucleons removed from the projectile by the ablation-abrasion
process. If the pre-fragment is in a thermal equilibrium, $T$
will be given by $T=\sqrt{E^{*}/a}$ and $a=A/k$, $k$ is the inverse level
density parameter. This leads to $T=\sqrt{13.3k\Delta A/A}$.
$\Delta A$ values are similar for the reactions studied here. Therefore $T$
becomes higher for the lighter pre-fragments which are produced more from the
lighter projectile. However one should note that the extracted temperatures
by the IYR from the cold fragments are significantly modified from that of the
pre-fragments by the sequential decay process. The correlation in Fig. \ref{Tandmu}(a)
merely reflects that the difference of the pre-fragments still sustains in some
extent through the sequential decay process.

\section{Summary}

In summary, the temperature of a fragment after sequential decay is
studied using the isobaric yield ratio method in the framework of a modified
Fisher model. The difference between the binding energy of the $I\geq5$ isobars
at zero $T$ [$\Delta B(0)$] is found to be valid for
substituting the value of $\Delta B(T)$ at low $T$. Relatively low $T$ which
range from 1 to 3MeV are obtained in different reactions. It is shown
that $T$ depends on the neutron-richness of the projectile. The mass of the target
used affects $T$ only slightly. The incident energy is found to affect $T$
very little. $\Delta\mu$ is found to depend linearly on the $N/Z$ of projectile,
i.e., larger $\Delta\mu$ is found in reactions induced by more neutron-rich
projectile. Due to the mass of the isobars of $A<50$,
an attempt was made to use the temperature-corrected $\Delta B(T)^{*}$.
It is found that $T$ of the $^{48}$Ca reaction, in which isobars of $A<50$
are dominant, is largely modified by the temperature-corrected
$\Delta B(T)^{*}$, while $T$ of other reactions, which involve isobars
of larger masses, only are slightly affected by the $\Delta B(T)^{*}$.

The SAA model analysis for the 140$A$ MeV $^{64}$Ni+$^9$Be
and 1$A$ GeV $^{136}$Xe+$^{208}$Pb reactions revealed that the secondary
decay process significantly modifies $T$ and $\Delta\mu$ of the pre-fragments
and governs those obtained from the cold fragments. This leads to
similar IYR distributions and temperature of cold fragment in reactions
of different incident energies and different masses of projectiles.
The SAA results also suggest that $T$ from IYRs indeed
reflects the actual physical temperature for the pre-fragments and
the final fragments, although the latter should be viewed as a
sequenced temperature of the secondary decay process in conjunction
with that of the primary process. Since the MFM method should be
applied to the initial hot nuclear matter in principle, the
drawback of the application of this method to the final fragments is
not directly to probe the characteristic nature of the initial stage,
but mainly to probe that of the secondary decay processes. However,
we believe that to elucidate the effect of the secondary process on
the fragment yield is crucial to studying the nature of the primary
emitting source, because all experimentally observed fragments have
to go through this process.

\begin{acknowledgments}
This work is supported by National Natural Science
Foundation of China under Contracts No. 10905017 and No. 10979074, the Knowledge
Innovation Project of the Chinese Academy of Sciences under Contract
No. KJCX2-EW-N01, the Program for Innovative Research Team (in Science and Technology)
under contract No. 2010IRTSTHN002 in Universities of Henan Province,
and the Young Teacher Project in Henan Normal University, China. One of us (R. Wada) thanks
"the visiting professorship of senior international scientists" of the Chinese Academy
of Sciences for support.
\end{acknowledgments}

\end{document}